\documentclass[fleqn,10pt]{wlscirep}
\usepackage[utf8]{inputenc}
\usepackage[T1]{fontenc}
\usepackage{subfig}
\usepackage{lineno}
\usepackage{float}
\usepackage{afterpage}

\usepackage{soul}
\usepackage{bbm}

\title{Controlling Dynamical Systems into Unseen Target States Using Machine Learning}

\author[1,*]{Daniel K{\"o}glmayr}
\author[2]{Alexander Haluszczynski}
\author[3]{Christoph R{\"a}th}
\affil[1]{Institut f{\"u}r KI-Sicherheit,
Deutsches Zentrum f{\"u}r Luft- und Raumfahrt (DLR), Ulm, Germany}
\affil[2]{risklab,
Allianz Global Investors,
Munich, Germany}
\affil[3]{Institut für Frontier Materials auf der Erde und im Weltraum, Deutsches Zentrum f{\"u}r Luft- und Raumfahrt (DLR), 51140 Cologne, Germany}

\affil[*]{daniel.koeglmayr@dlr.de}

\begin{abstract}

We present a novel, model-free, and data-driven methodology for controlling complex dynamical systems into previously unseen target states, including those with significantly different and complex dynamics. Leveraging a parameter-aware realization of next-generation reservoir computing (NGRC), our approach accurately predicts system behavior in unobserved parameter regimes, enabling control over transitions to arbitrary target states utilizing a new prediction evaluation and selection scheme. Crucially, this includes states with dynamics that differ fundamentally from known regimes, such as shifts from periodic to intermittent or chaotic behavior. The method's parameter awareness facilitates non-stationary control with which control scenarios are generated and evaluated on the basis of predefined control objective. In addition to proving the method for transient-free control to extrapolated chaotic target states over transition times, we demonstrate the method's effectiveness on a nonlinear power system model. Our method successfully navigates transitions even in scenarios where system collapse is observed frequently, while ensuring fast transitions and avoiding prolonged transient behavior. By extending the applicability of machine learning-based control mechanisms to previously inaccessible target dynamics, the methodology opens the door to new control applications while maintaining  exceptional efficiency.

\end{abstract}
\begin{document}

\flushbottom
\maketitle

\thispagestyle{empty}

\section*{Introduction}
Controlling dynamical systems is a fundamental challenge in engineering and science, crucial for maintaining stability, optimizing performance, and enabling adaptability across a broad spectrum of applications. From aerospace engineering, where precise control of aircraft and spacecraft is vital \cite{stevens2015aircraft,crouch1984spacecraft}, to industrial automation, where maintaining efficiency and safety in manufacturing processes is essential \cite{leitao2009agent}, the ability to effectively control dynamical systems is driving significant advancements among different fields \cite{qin2000overview,niku2020introduction,kim2019unmanned,ding2018model,liu2011controllability}. In biomedical engineering, controlling physiological systems and biomedical devices improved patient care\cite{dey2024advances,krauss2021technology,hochberg2012reach}. In the realm of energy systems, it enabled the efficient operation of power grids and the integration and optimization of renewable energy sources \cite{camacho2012control,apata2020overview}. However, the increasing integration of distributed renewable energy sources poses challenges to the robustness of power grids. Consequently, it is imperative to develop new control strategies tailored for future micro grids to ensure stability and reliability \cite{smith2022effect}. Traditionally, control methods such as Proportional-Integral-Derivative (PID) controllers, state-space control, and model predictive control (MPC) have been employed to manage these systems. While effective in many cases, these approaches often rely on accurate mathematical models and can struggle with the complexities inherent in nonlinear, high-dimensional, and time-varying systems \cite{iqbal2017nonlinear, alora2023practical, chen2024learning}. In recent years, the advent of machine learning introduced powerful new tools for predicting and controlling dynamical systems. Techniques such as deep learning, neural ordinary differential equations and reinforcement learning showed promise in modeling complex systems and developing control strategies \cite{mnih2015human,bottcher2022ai,degrave2022magnetic}. However, these methods often require extensive computational resources and can be difficult to interpret and implement. Reservoir computing (RC) emerges as a promising candidate in this landscape, offering a simpler yet effective approach for handling the temporal dependencies and nonlinearities of dynamical systems. Reservoir computing leverages a fixed, high-dimensional dynamical system, the reservoir, to process temporal data. The key advantage of reservoir computing lies in its separation of dynamic processing, managed by the reservoir, from the training process, which optimizes only the readout layer. This significantly reduces the computing costs and makes it orders of magnitude more efficient than comparable algorithms\cite{shahi2022prediction,herteux2024forecasting}.
Recent advancements in reservoir computing led to the development of various reservoir architectures, enhancing their predictive capabilities and computational efficiency. Traditional reservoir computing employs a randomly generated network at its core \cite{jaeger2004harnessing}, providing robust performance in learning and predicting temporal patterns. Minimal reservoir computing simplifies this approach significantly by using a block-diagonal matrix structure, which lowers computational complexity and minimizes training data requirements \cite{ma2023novel}. Additionally, the next-generation reservoir (NGRC) computing architecture, originating from nonlinear vector autoregression, utilizes a deterministic library of unique monomials derived from the input data and time shifts of it \cite{gauthier2021next}. Studies comparing these architectures demonstrated their effectiveness across various applications. For instance, the prediction performance of these architectures on chaotic systems showed that minimal reservoir computing and next-generation reservoir computing outperforms traditional methods in scenarios with limited data availability, providing more accurate and reliable predictions \cite{ma2023novel}. Further research showed that next-generation reservoir computing can achieve comparable predictive performance to the already data-efficient traditional reservoir computing architectures while requiring only a fraction of the training data, i.e. about $10$ to $100$ times less \cite{gauthier2021next,barbosa2022learning,haluszczynski2023controlling}. These innovations make reservoir computing as a framework more flexible and effective, especially in scenarios with limited data and strong nonlinearity, opening up its use in various fields and problems. One field is the control of dynamical systems \cite{haluszczynski2021controlling,haluszczynski2023controlling,kent2024controlling,kent2024controlling_}. In \cite{haluszczynski2021controlling}, traditional reservoir computing is used to control dynamical systems into arbitrary periodic, intermittent or chaotic target states without requiring knowledge of the underlying governing equations. To achieve effective control, the reservoir computer must accurately capture the statistical behavior of the target dynamics. Subsequently, a control force—derived from the RC's prediction and the current system state—is applied to control the system toward the desired target state. In \cite{haluszczynski2023controlling}, the performance of this approach was compared with one using NGRC at its core. Although both methods perform similarly in reproducing the statistical climate of the target states during control, NGRC outperformed the RC approach by requiring roughly ten times less training data. 
In this work, we now make use of the extrapolative capabilities of parameter-aware NGRC by learning a dynamical digital twin of the dynamical systems analyzed from few parametrized data sets\cite{koglmayr2024extrapolating}. We introduce a general prediction evaluation and selection scheme to generate control strategies that fulfill a predefined control objective, with which the dynamical systems are controlled in extrapolated and previously unseen target states. Thereby we demonstrate the use of the introduced control methodology in critical control scenarios, where previous control approaches would lead to transient behavior or even system collapse.

\section*{Results}

We evaluated the proposed control method on two dynamical systems with distinct objectives and application contexts. First, we control the Lorenz system from an initially periodic orbit to an extrapolated weakly chaotic state, imposing the requirement that no chaotic transients arise during the parameter transition, a phenomenon that is frequently observed in simulations of this parameter change but is consistently suppressed by our approach across a wide range of transition times. Second, we apply the controller to a power system model whose dynamics is governed by reactive power demand and are prone to rich chaos and even voltage collapse. Starting in a periodic regime, we execute an instantaneous parameter change to an extrapolated periodic window located just before the collapse threshold, achieving fast transitions to the target state while reliably preventing both collapse and the prolonged chaotic transients typically observed in simulations. In both studies, we train a parameter-aware NGRC on few parameterized data sets, allowing it to serve as a dynamical digital twin that efficiently generates non-stationary candidate trajectories toward extrapolated targets, which are evaluated and selected if they satisfy a desired control objective to become the control strategy. For the Lorenz system and its weakly chaotic target state, we utilize a running correlation dimension metric for the evaluation scheme. For the power system model and its initial periodic state and periodic target state, we evaluate the control scenarios generated by measuring the spectral entropy over time with the objective of minimizing the transition duration.

\afterpage{
    \begin{figure}[!t]
    \centering
    \begin{minipage}{\linewidth}
      \centering
      \includegraphics[width=1\linewidth]{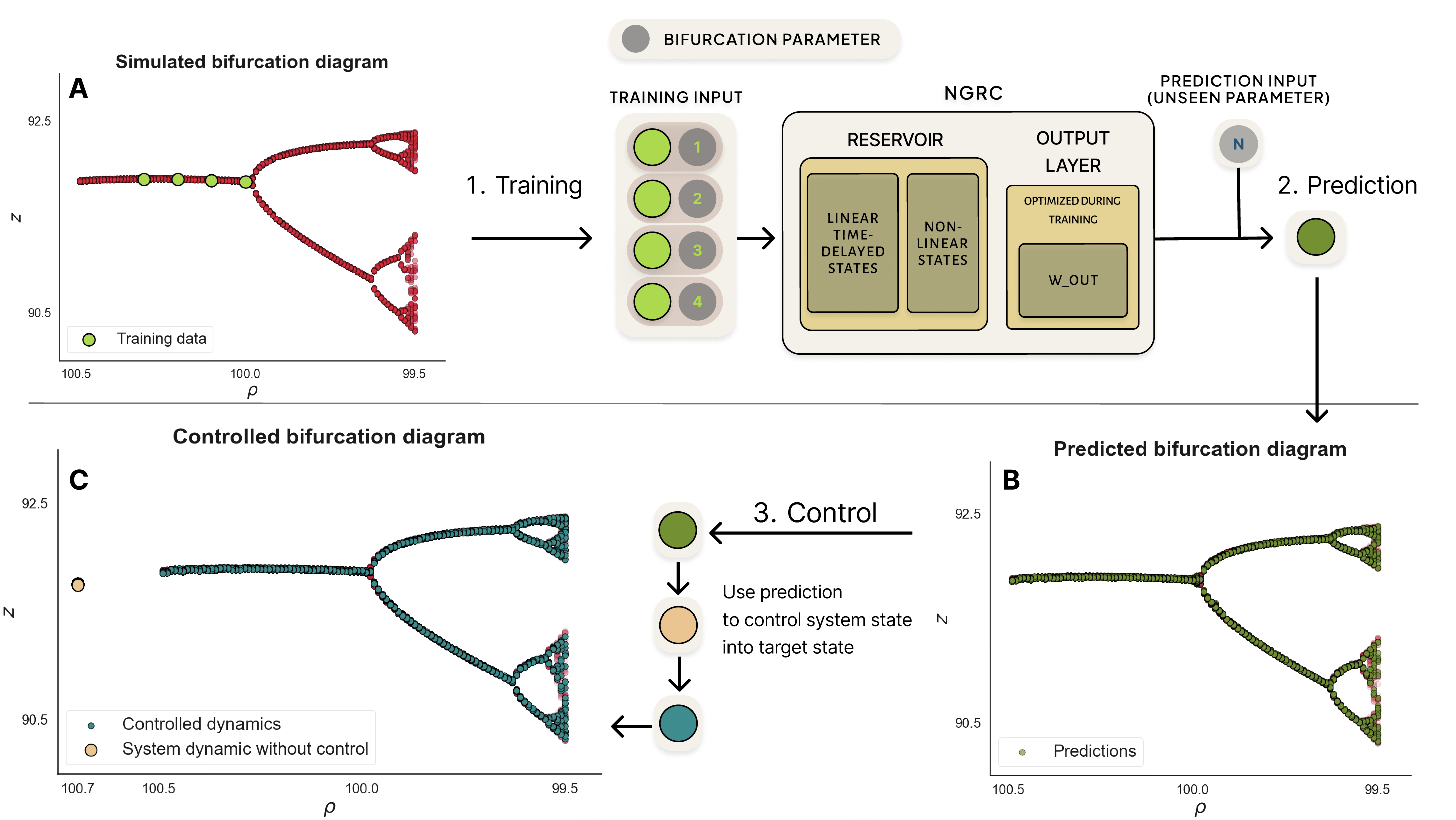}
      \caption{$\textbf{Controlling dynamics into unseen states.}$
      $\textbf{A}$) Segments of the Lorenz bifurcation diagram (red). Four training samples and their bifurcation parameters are used for training (light green). $\textbf{B}$) Predicted bifurcation diagram for unseen parameter regions using the trained model. $\textbf{C}$) Controlled bifurcation diagram demonstrates the ability of the control mechanism to control system states into arbitrary and unseen target states using the model’s predictions.}
      \label{fig:turbulence_probability_prediction}
    \end{minipage}
    
    \vspace{2em} 
    
    \begin{minipage}{\linewidth}
      \centering
      \begin{minipage}[b]{0.49\textwidth}
        \centering
        \includegraphics[width=\textwidth]{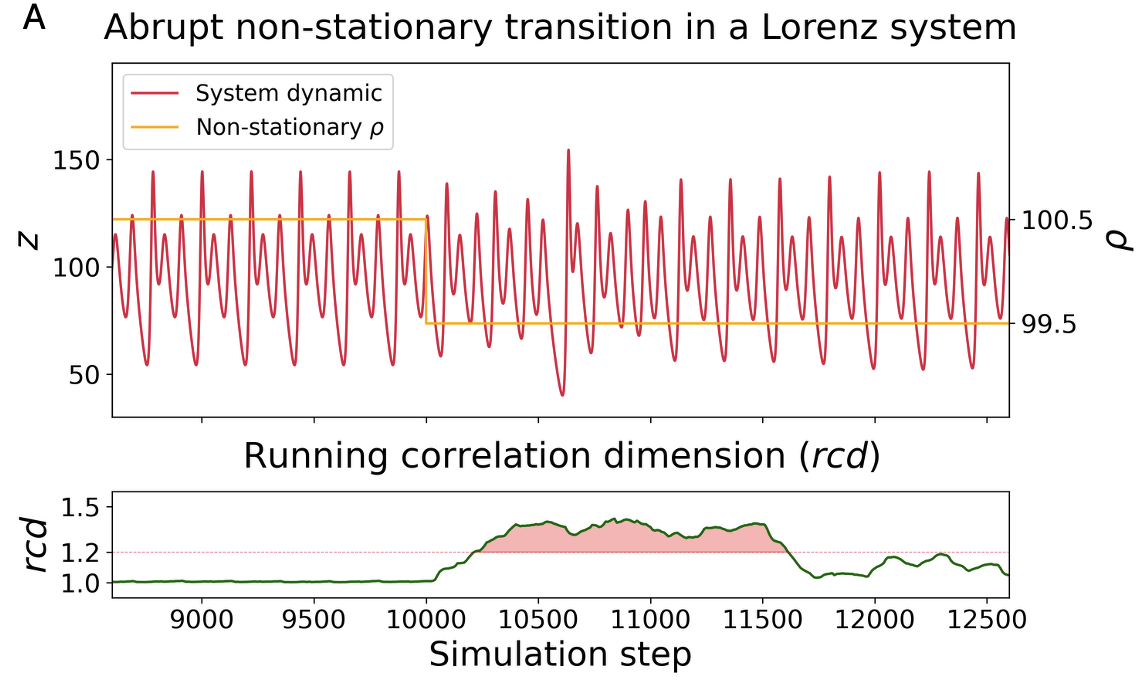}
      \end{minipage}
      \hfill
      \begin{minipage}[b]{0.49\textwidth}
        \centering
        \includegraphics[width=\textwidth]{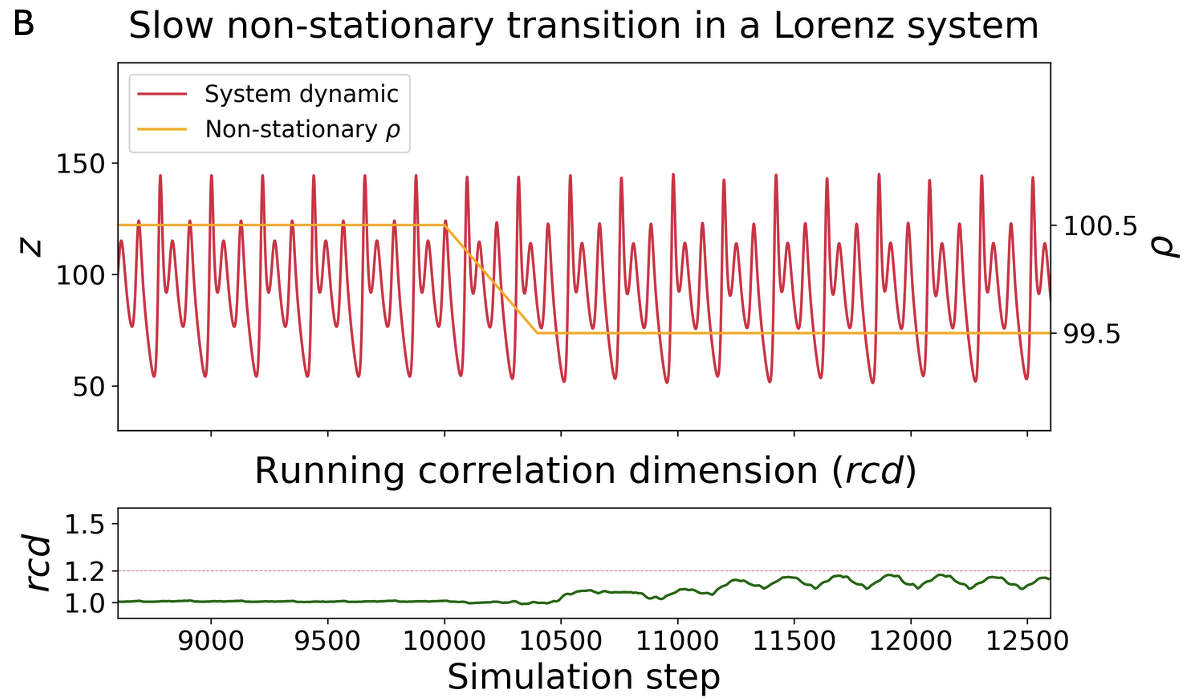}
      \end{minipage}
      \caption{$\textbf{Non-stationary transition duration influences system behavior.}$ 
      The orange line \textbf{A}  and \textbf{B} shows the non-stationary parameter transition over time, while the red lines represent the corresponding system dynamics from simulations. In \textbf{A}, an abrupt parameter transition causes transient behavior after the transition, which is measured by the increased running correlation dimension (\textbf{A}, green). In \textbf{B}, a slow parameter transition leads to a smooth progression toward the target dynamics.}
      \label{fig:turbulence}
    \end{minipage}
    \end{figure}
}

\subsection*{Lorenz system}
We test the method on the Lorenz system by varying the parameter $\rho$ in the vicinity of $\rho=100$. For $\rho > 100$, the system exhibits periodic behavior. As $\rho$ decreases below 100, it undergoes period-doubling bifurcations and transitions to weakly chaotic dynamics. The bifurcation behavior is colored red in Fig. \ref{fig:turbulence_probability_prediction} $\mathbf{A}$ and has previously been analyzed for the prediction of unseen states using traditional reservoir computing \cite{kim2021teaching}.
We use parameter-aware NGRC \cite{koglmayr2024extrapolating} to learn the Lorenz system using four periodic and stationary training data samples with parameters $\rho_{\text{train}}=100.0, 100.1, 100.2$ and $100.3$. After successful training, the method can extrapolate and interpolate the behavior of the system in the unseen parameter regions of $\rho$, as shown in Fig. \ref{fig:turbulence_probability_prediction} $\mathbf{B}$. For each parameter, a total data length of $5.000$ time steps is sufficient to train the NGRC. We use the trained NGRC as the basis for the control approach and leverage its dynamical similarity to the actual dynamical system to generate scenarios for the control strategy. First, we test for the simple case of stationary target states as a control objective, inter- and extrapolated by our parameter-aware approach. A control force is applied, defined as the product of a control parameter $K$ and the difference between the current state of the system at $\rho_{\text{test}}$ and the predicted state of the target dynamics at $\rho_{\text{pred}}$. The control force directs the system's dynamics at each time step to converge with the predicted dynamics. For example, the dynamic at $\rho_{\text{test}} = 100.7$ can be controlled to align with any $\rho$ value within the range of the analyzed bifurcation diagram, as shown in Fig. \ref{fig:turbulence_probability_prediction} $\mathbf{C}$.

\begin{figure}[htbp]
\centering
\includegraphics[width=1\linewidth]{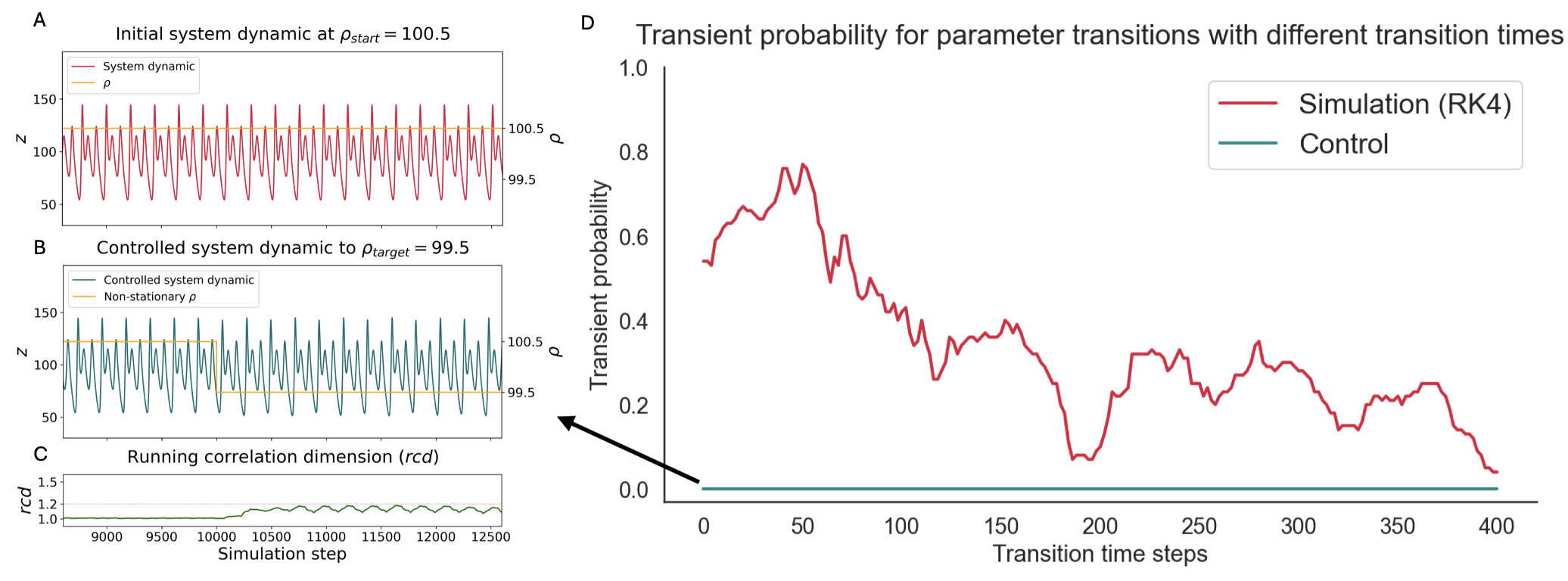}
\caption{$\textbf{Control method facilitates smooth transitions to unseen target states.}$ 
The Lorenz system is controlled from a periodic initial state at $\rho_{\text{start}} = 100.5$ to a weakly chaotic state at $\rho_{\text{target}} = 99.5$, using linearly decreasing parameterizations over various transition durations, shown in \textbf{D}. The control method enables a smooth, transient-free transition to the target state (\textbf{D}, blue). In contrast, simulations with the same parameter transitions (\textbf{D}, red) show that shorter transition durations often cause transients. For each transition time step, 100 differently initialized system were tested. \textbf{A} illustrates one of those initial periodic trajectory at $\rho_{\text{start}}$ (\textbf{A}, red). \textbf{B} and \textbf{C} show the controlled dynamics to the weakly chaotic target state with an instantaneous parameter change (\textbf{B}, blue) and its running correlation dimension (\textbf{C}, green).}
\label{fig:turbulence_props}
\end{figure}

\begin{figure}[!t]
  \centering
  \begin{minipage}[b]{0.45\textwidth}
    \centering
    \includegraphics[width=\textwidth]{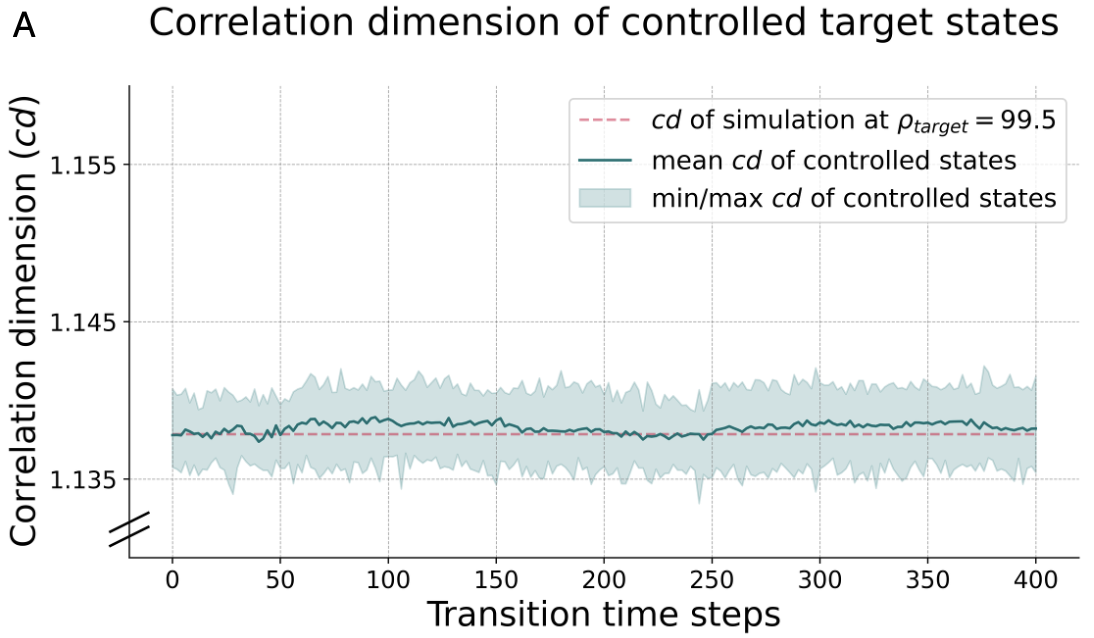} 
  \end{minipage}
  \hfill  
  \begin{minipage}[b]{0.45\textwidth}
    \centering
    \includegraphics[width=\textwidth]{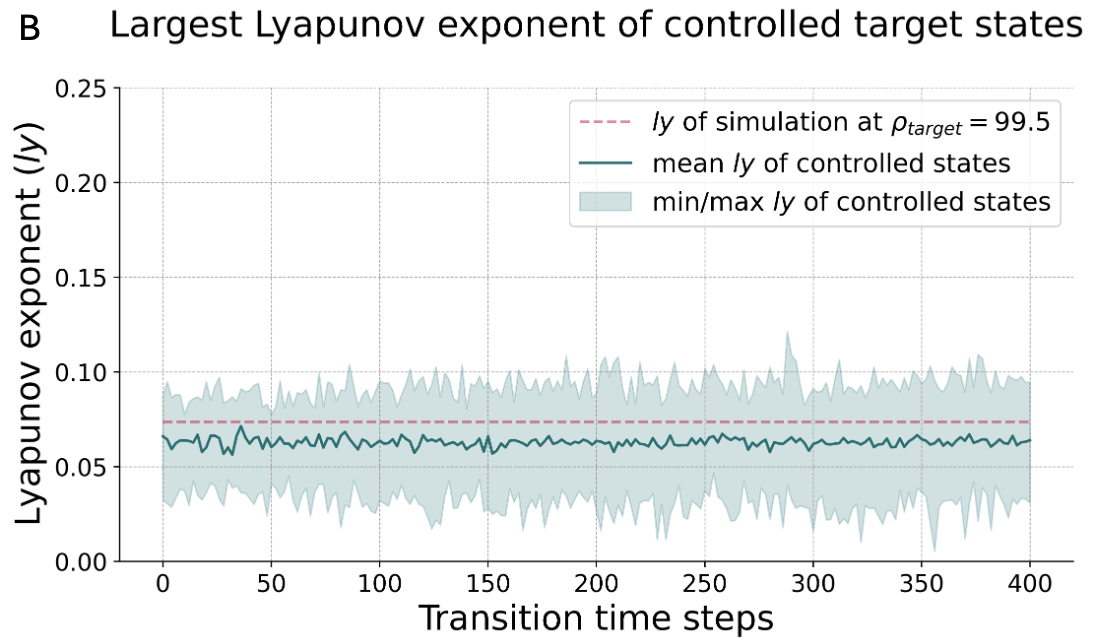} 
  \end{minipage}
  \caption{$\textbf{Statistical climate of controlled target states.}$ 
Evaluation of the statistical climate of the $100$ controlled Lorenz systems at $\rho_{target} = 99.5$ given each transition period shown in Fig. \ref{fig:turbulence_props}. In \textbf{A}, the correlation dimension is measured and in the \textbf{B} the largest Lyapunov exponent.
}
  \label{fig:statistical_climate}
\end{figure}

Subsequently, we test the control approach for non-stationary dynamics. As shown in Fig. \ref{fig:turbulence} $\mathbf{A}$, abrupt changes in $\rho$ (e.g. from $100.5$ to $99.5$) can lead to transient behavior in the system dynamics post-transition \cite{manneville1979intermittency}, whereas the gradual parameter change in Fig. \ref{fig:turbulence} $\mathbf{B}$ results in a smooth transition. 
This behavior is further analyzed in the right plot of Fig. \ref{fig:turbulence_props}, where we analyze how the system simulations behave when they transition from a periodic orbit at $\rho_{\text{start}}=100.5$ to a weakly chaotic regime at $\rho_{\text{target}}=99.5$. Here, the transitions are performed with linearly changing $\rho$ values. Depending on the transition time and the location of the trajectory when the parameter transition starts, transient behavior might occur. We capture the transient behavior with a running correlation dimension (rcd) \cite{lehnertz1998can,grassberger1983generalized} approach (detailed in Methods), by defining that if the rcd becomes larger during the transition than that of the weakly chaotic target state, transient behavior occurred. For each analyzed transition time, we apply this measure on 100 randomly initialized Lorenz systems to define a probability of how likely it is for the transition time that the system exhibiting transient dynamics. Now, we leverage this measure to generate control strategies for each transition time to avoid transients in the system during the parameter change. Therefore we employ the parameter-aware NGRC to predict from different initial conditions non-stationary control scenarios, inducing transitional periods from $\rho_{\text{start}}$ through intermediate $\rho$ values to the extrapolated target state at $\rho_{\text{target}}$, and select those scenarios, where the rcd stays below the extrapolated maximum of the weakly chaotic target state during the transition as the control strategy. With this approach we are able to reduce the transient probability for all analyzed transition periods quantitatively to zero, shown in Fig. \ref{fig:turbulence_props} $\mathbf{D}$. In Fig. \ref{fig:turbulence_props} $\mathbf{A}$ the initial system dynamic of the Lorenz system at $\rho_{\text{start}}$ is shown, together with a controlled trajectories for an abrupt transition (Fig. \ref{fig:turbulence_props} $\mathbf{B}$) and its rcd (Fig. \ref{fig:turbulence_props} $\mathbf{C}$). The weakly chaotic state is visible in changes in amplitudes of the oscillations after the parameter change. We evaluate the quality of the controlled dynamics by comparing the statistical climate of the actual system at $\rho_{\text{target}}$ to that of the $100$ controlled systems for each transition period. Therefore we measure the spatial complexity as well as the temporal complexity for each controlled system by calculating its correlation dimension\cite{grassberger1983generalized} and its largest Lyapunov exponent\cite{rosenstein1993practical}, respectively. For the actual system and each controlled system, the measurements were carried out over $25.000$ time steps and are displayed in Fig. \ref{fig:statistical_climate}. The correlation dimension of the actual system is $cd = 1.1378 $, reproduced with an average correlation dimension of $cd = 1.1382$ across all controlled systems. Similarly, the largest Lyapunov exponent of the actual system is $ly = 0.0735$, while the average largest Lyapunov exponent across all controlled systems is $ly = 0.0628$. Both metrics, particularly the correlation dimension, show that the statistical climate of the actual system is effectively reproduced following each transition period, despite the parameter-aware NGRC being trained exclusively on periodic trajectories and is extrapolating the emerging chaotic system behavior. With that the method enables smooth control into unseen target states preventing adverse transient behavior during transitions.

\subsection*{Power system model}

\begin{figure}[htbp]
\centering
\includegraphics[width=\linewidth]{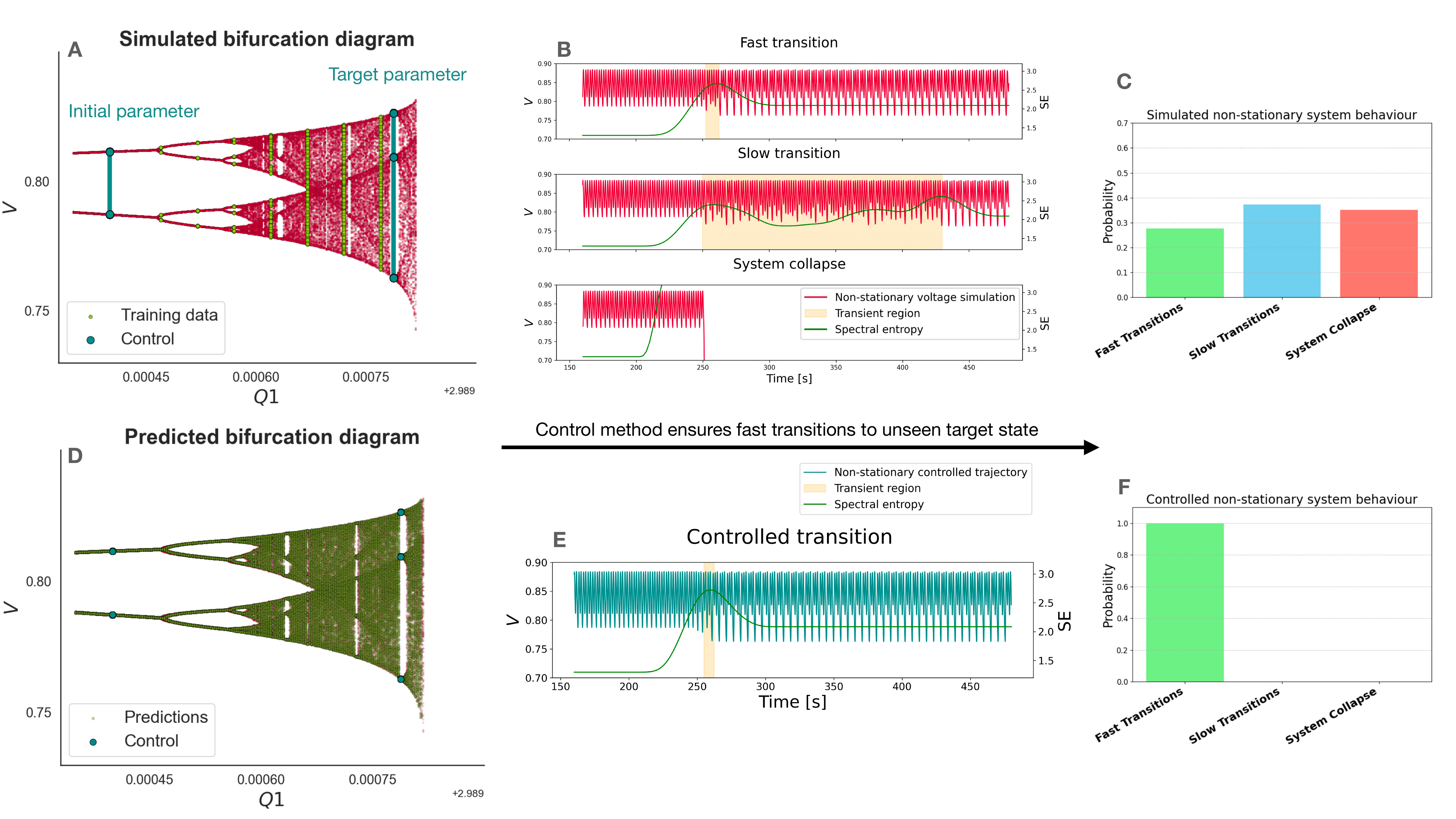}
\caption{$\textbf{Control method applied to power system control.}$
Fig. \textbf{A} illustrates the voltage behavior of the power system model for several reactive power demands $Q_1$. The NGRC is trained on seven training data samples (shown in light green), to accurately interpolate and extrapolate the system behavior in unseen parameter regions of $Q_1$ (shown in Fig. \textbf{D}).
We test the methods control capabilities by applying an instantaneous parameter switch from an unseen initial regime to an unseen target regime, both of which are periodic (Fig. \textbf{A} cyan). From simulating this switch, three different behaviors occur (Fig. \textbf{B}) with their probability for 250 different initial condition shown in Fig. \textbf{C}. The control method applied to the initial system dynamic, ensures fast transitions to the target dynamic while completely avoid prolonged chaotic transients and system collapses (shown in Fig. \textbf{E} and Fig. \textbf{F}).}
\label{fig:power_system}
\end{figure}

\begin{figure}[htbp]
  \centering
  \begin{minipage}[b]{0.435\textwidth}
    \centering
    \includegraphics[width=\textwidth]{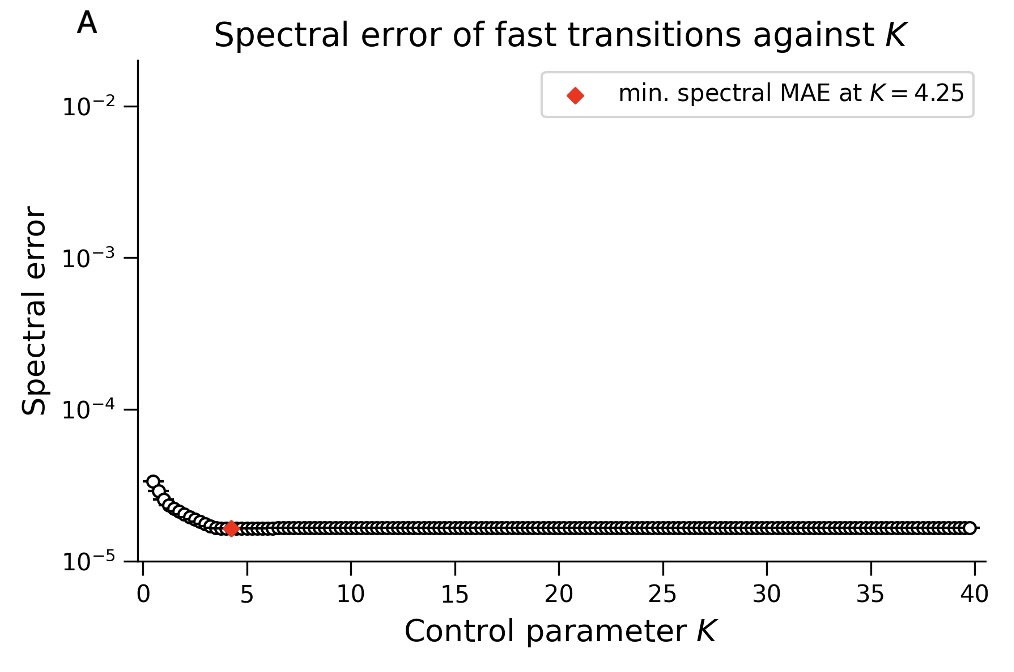} 
  \end{minipage}
  \hfill  
  \begin{minipage}[b]{0.485\textwidth}
    \centering
    \includegraphics[width=\textwidth]{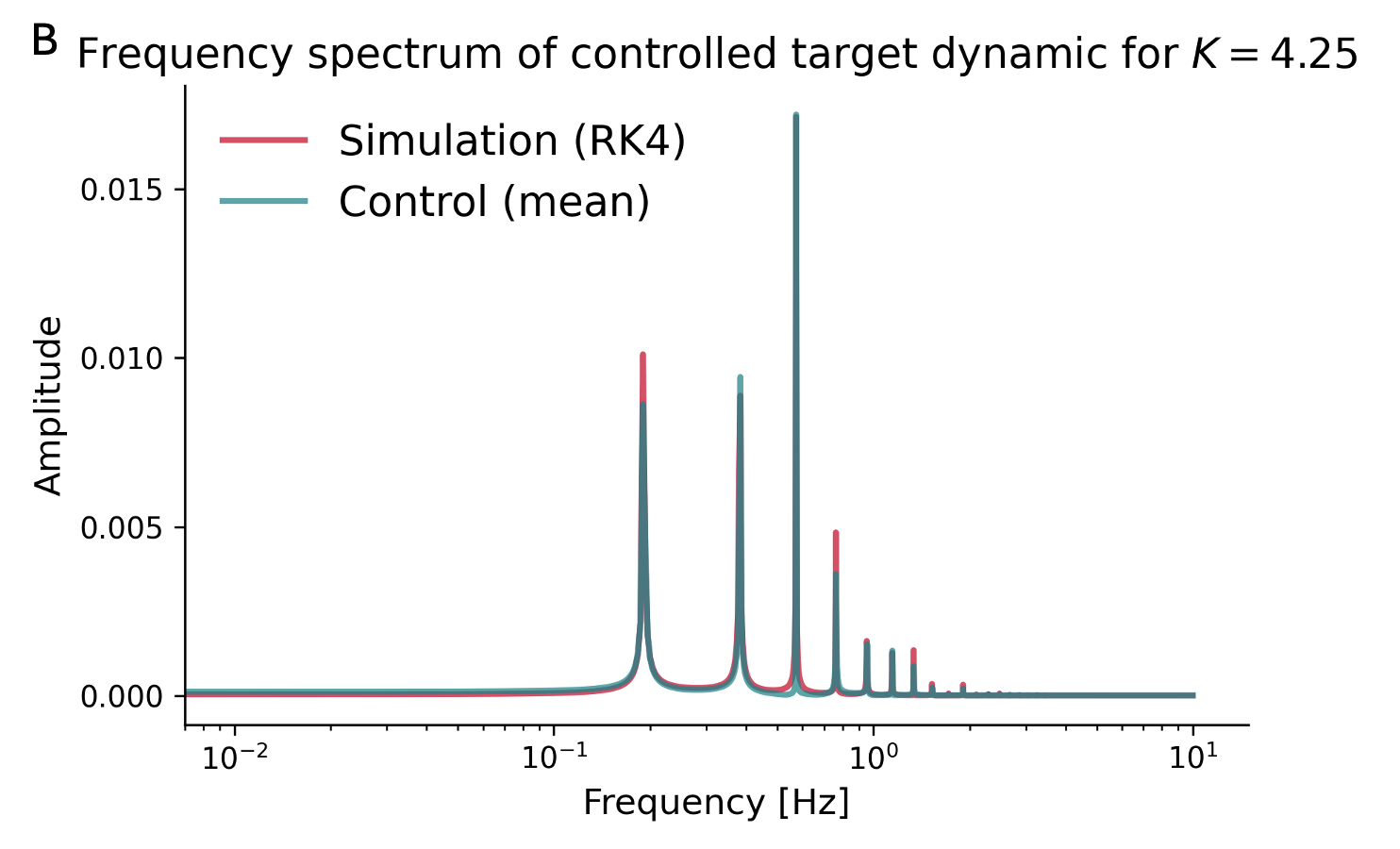} 
  \end{minipage}
  \caption{$\textbf{Accuracy of the controlled and extrapolated target dynamics.}$ To quantify the accuracy the controlled dynamics in the extrapolated target state, we calculate the mean absolute error of the frequency spectra (Spectral Error) of the simulated and controlled dynamics at the target parametrization. \textbf{A} shows the spectral error of the 250 controlled trajectories as box plots for different control parameters. \textbf{B} shows the mean frequency spectrum of 250 controlled trajectories for the control parameter associated with the smallest spectral error and the target spectrum from simulation.}
  \label{fig:noise_free}
\end{figure}

In this section, we apply the control mechanism on a generic power system model introduced by Dobson \emph{et al.}~\cite{dobson1989towards} that models the system collapse in small power systems (see Methods). We take the reactive power demand $Q_1$ as the bifurcation parameter and train the parameter-aware NGRC with 7 differently parametrized datasets in $Q_1$ (Fig.\ref{fig:power_system} \textbf{A}). With that we can interpolate and extrapolate the system behavior in unseen parameter regimes and find a periodic window in the vicinity of a system collapse (Fig. \ref{fig:power_system} \textbf{D}). We take the associated parameter $Q^{window}_1=2.989788$ as the target regime $Q^{target}_1$ to where we want to control the system to from a periodic state at $Q^{inital}_1 = 2.98940$, far from the control target with rich chaotic behaviors between the parameter range.
By analyzing this parameter change through an instantaneous parameter change, using non-stationary simulations with the Runge-Kutta 4 method and 250 different initial conditions, we observe that $35\%$ of the trajectories result in a system collapse (Fig. \ref{fig:power_system} \textbf{B}), even though the parameterization has not yet reached the critical values typically associated with a system collapse event.
Since the system exhibits rich chaotic behaviors between the initial and target parameterizations, $37\%$ of the simulations show trajectories that transition into the periodic target regime with prolonged transient behavior (slow transition). These transitions are characterized by transient periods lasting longer than $12.5$ seconds. The remaining $28\%$ of simulations exhibit fast transitions with transient periods shorter or equal $12.5$ seconds (Fig. \ref{fig:power_system} \textbf{C}).
To quantify the system behavior, we calculated the spectral entropy (see Methods) of the trajectories, which remains constant for periodic dynamics. Since the target dynamic is also periodic, we measure the transient periods as the time required for the spectral entropy to reach a constant value again, adjusted by the overlap used in the short-time fourier transformation domain. The spectral entropy for different system behaviors is shown in Fig. \ref{fig:power_system} \textbf{B} and \textbf{E} in green and the associated detected transient regions in orange. 
We apply the control method to control the power system model from the initial periodic regime at $Q^{inital}_1$
to the extrapolated periodic window in the vicinity of the system collapse at $Q^{target}_1$, under the objective to avoid system collapse and prolonged transient transitions observed for the non-stationary simulations. 
Therefore, we predict the non-stationary dynamics of the parameter switch from varying initial conditions with the trained NGRC until we find a trajectory that fulfills the control objective. This trajectory is than used to define the control forces applied to each of the 250 previous initial systems, remaining at  $Q^{inital}_1$ to control them into the target regime at $Q^{target}_1$. In Fig. \ref{fig:power_system} \textbf{E} one of the controlled voltage dynamics with the spectral entropy measure and the detected transient region is displayed. In Fig. \ref{fig:power_system} \textbf{F} we  measured for each of the 250 previous initial systems the transient durations when controlled from their initial parameterization to the target parameterization and find that the control method ensures quantitatively the objective fast transitions, even for initial power system dynamics which would exhibit a system collapse or prolonged transit dynamics. 
To qualitatively evaluate the controlled dynamics at the target state, we compare their frequency spectrum with that of the actual voltage dynamics at $Q^{target}_1$ and define the spectral error as the mean absolute error between the two spectra. In Fig. \ref{fig:noise_free} $\mathbf{A}$ the spectral error is plotted for different control parameters $K$ ranging from 0.25 to 40.
For small control parameters this error decreases until a minimum at $K_{min}=4.25$ is reached and then slightly increases until the control method breaks down for control parameters equal or larger than 40, where the dynamics exhibiting a system collapse. The frequency spectrum of the controlled dynamics for $K_{min}$ and the ground truth is plotted in Fig. \ref{fig:noise_free} $\mathbf{B}$, confirming that the NGRC controller is not only able to extrapolate the dynamics and the associated unseen frequencies accurately but also enables the accurate control to these states.

\newpage
\subsubsection*{Power system model with noise}

\begin{figure}[htbp]
\centering
\includegraphics[width=0.95\linewidth]{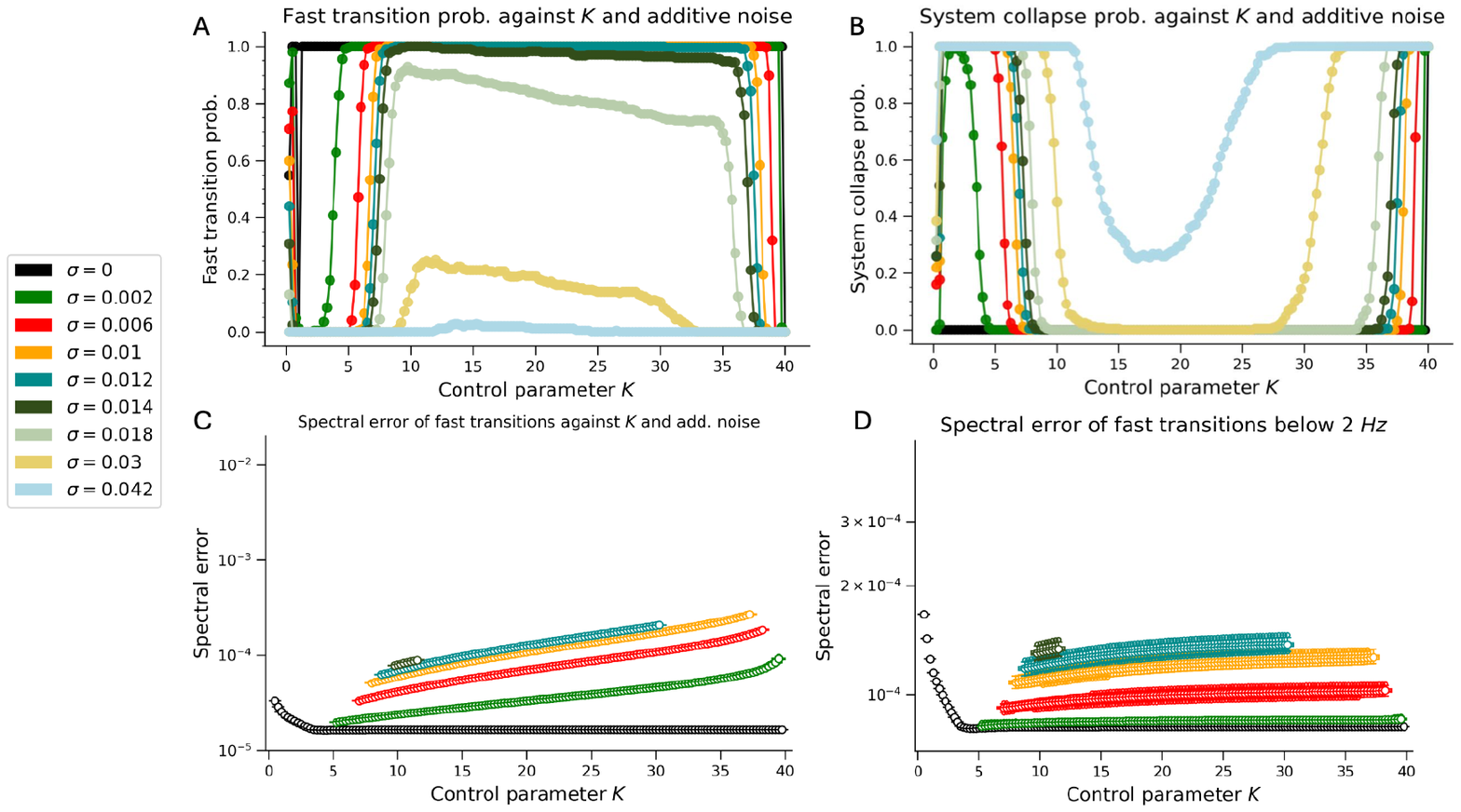}
\caption{$\textbf{Control with additive noise.}$ The upper plots show the system behavior of the controlled dynamics given different control parameters and noise levels. For each configuration we show the probability with which the 250 controlled dynamics exhibiting the objective fast transition to the target dynamic (\textbf{A}) and a system collapse (\textbf{B}). \textbf{C} shows the spectral error between the fast transitioned controlled target dynamics and the target dynamic without noise. \textbf{D} shows the corresponding error for frequencies below $2 Hz$, where the frequencies of the target dynamics are. The results highlight that on the one hand the control method can accurately control the initial dynamics to the target dynamics, even in high noise regimes with a sufficiently high and not higher control parameter $K$. On the other hand choosing an high $K$ leads to additional noise (mainly high frequency) by the control method itself.}
\label{fig:noise_control}
\end{figure}

\begin{figure}[htbp]
\centering
\includegraphics[width=\linewidth]{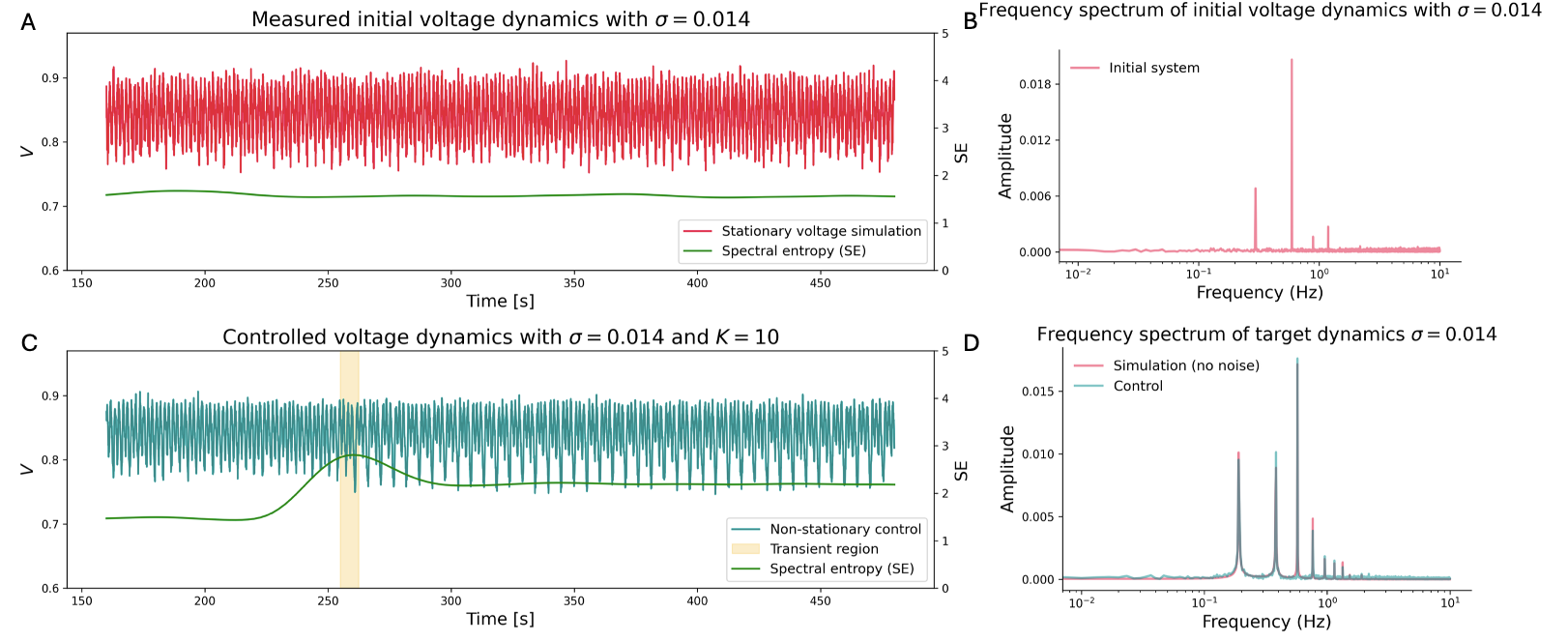}
\caption{$\textbf{Control with additive noise.}$ The upper plots show the voltage dynamics of the power system model with additive gaussian noise $\sigma = 0.014$ and its frequency spectrum. The NGRC controller processes the noisy input data and produces a force with control parameter $K=10$ and the differences between its predictions and the noisy input data to control the power system to the target state while maintaining the fast transition period objective (lower plots).}
\label{fig:noise14}
\end{figure}

\begin{figure}[htbp]
\centering
\includegraphics[width=\linewidth]{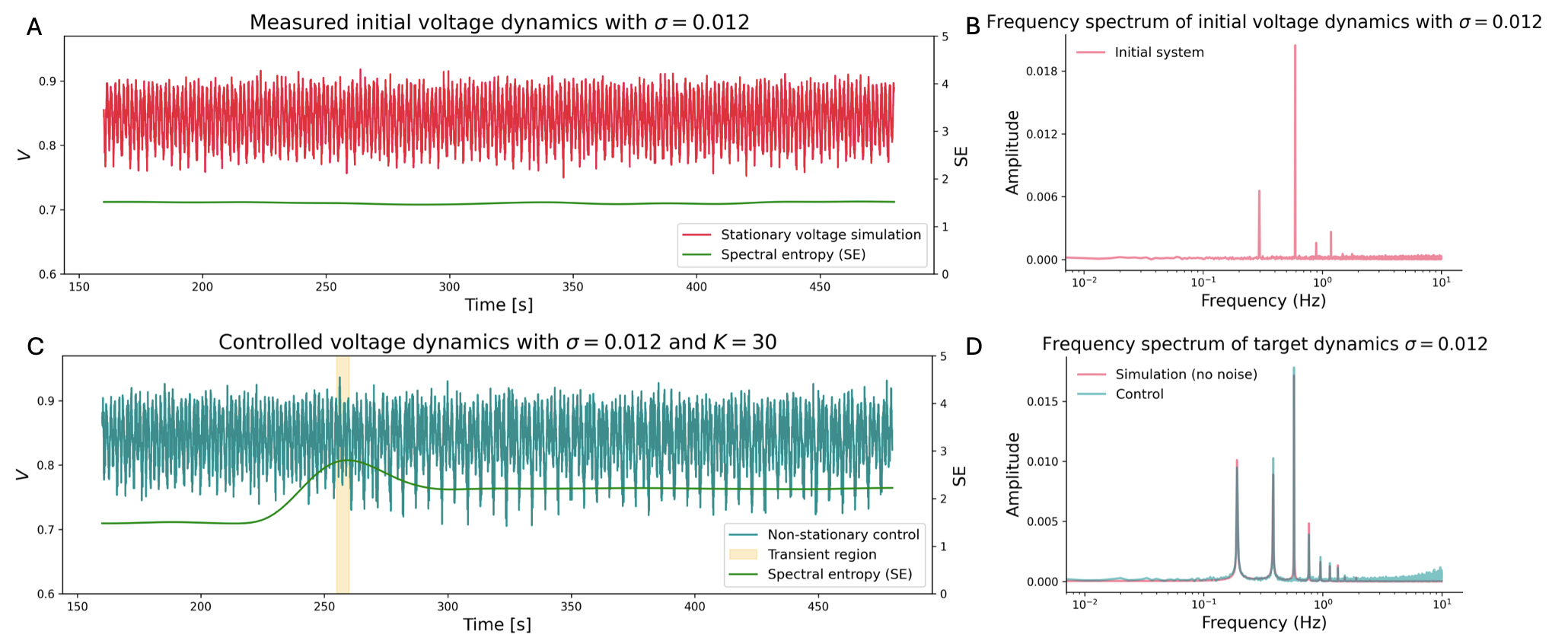}
\caption{$\textbf{Control with additive noise $\sigma=0.012$.}$ The upper plots show the voltage dynamics of the power system model with additive gaussian noise $\sigma = 0.012$ and its frequency spectrum. While the noise is slightly smaller than in Fig. \ref{fig:noise14}, the increased control parameter amplifies the noise in the controlled power system.}
\label{fig:noise12}
\end{figure}

\begin{figure}[htbp]
  \centering
  \begin{minipage}[b]{0.47\textwidth}
    \centering
    \includegraphics[width=\textwidth]{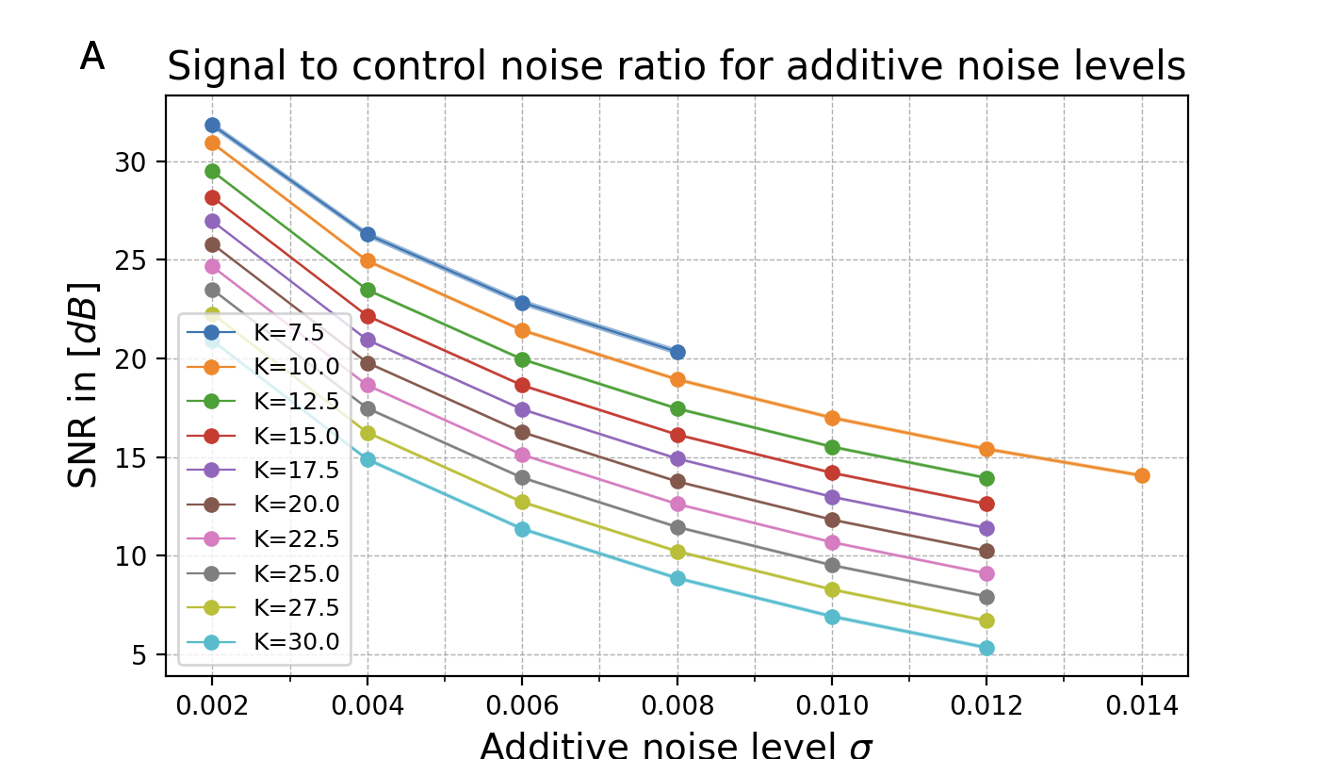} 
  \end{minipage}
  \hfill  
  \begin{minipage}[b]{0.47\textwidth}
    \centering
    \includegraphics[width=\textwidth]{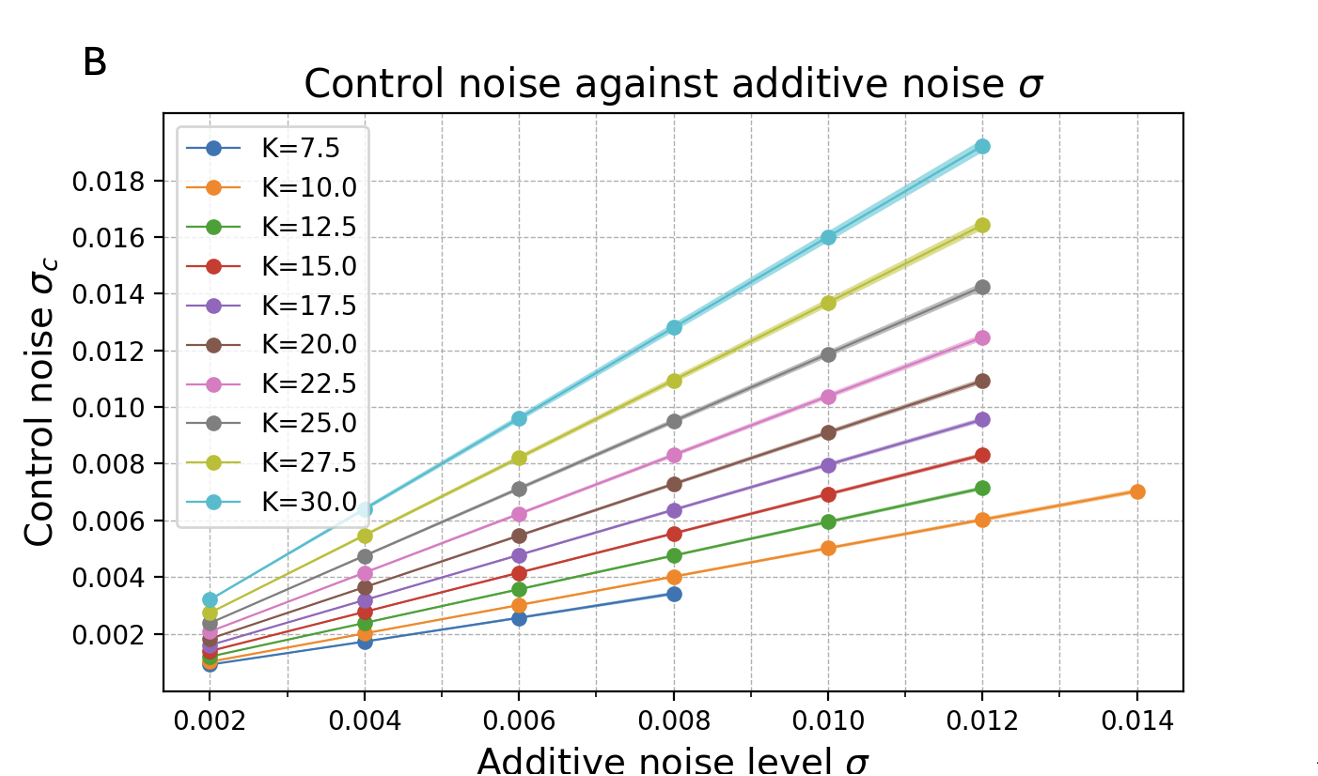} 
  \end{minipage}
  \caption{$\textbf{Noise induced through control.}$ 
\textbf{A} displays the mean signal to noise ratios for the $250$ fast transitioning controlled dynamics at the target dynamic with their standard deviation for different control parameters $K$ and additive noise levels $\sigma$. \textbf{B} shows the control noise on the controlled target dynamics originating from the control method itself for different additive noise levels and control parameters.
}
  \label{fig:noise_analyis_control}
\end{figure}

We test the performance of the control method for different levels of additional noise. In this setup the time series of the power system has additional gaussian noise with standard deviation $\sigma$ added to the system at each time step the system is evolving. We test the control objective of fast transitioning from $Q^{initial}_1$ to $Q^{target}_1$ for different $\sigma$ with the identical controller of the previous results. Only this time the controller processes noisy data $\tilde{\mathbf{u}}(t) = \mathbf{u}(t) + \mathbf{\mu}(t)$ with $\mathbf{\mu}(t)  \sim \mathbf{\mathcal{N}}(0,\,\sigma^{2})$ to calculate the control force, which is applied directly onto the (noise free) power system.
We measure the transient duration of the actual controlled power system with the spectral entropy. In Fig. \ref{fig:noise_control}, we test this setup for different control parameters $K$ and $\sigma$ ranging from 0.002 to 0.05, with 250 initial systems per configuration. While in the noise free case a well chosen control parameter might lead to a slightly reduced spectral error, it becomes increasingly important in the noise case in both the quantitative as well as qualitative metrics. Figure \ref{fig:noise_control} $\mathbf{A}$ shows the fast transition probabilities. The control objective of fast transitioning to the target dynamic is ensured up to $\sigma=0.014$ in a small range of control parameters in the neighborhood of $K=10$ (see Fig. \ref{fig:noise14} $\mathbf{C}$). For higher noise levels, the fast transition probability drops across all control parameters while the probability of a system collapse generally rises or becomes certain for small and high control parameters in the analyzed range for increasing noise levels.
For the noise and control parameter configurations which follow the control objective of ensuring fast transitions, we calculate the spectral error of the frequency spectrum of the controlled target states and the frequency spectrum of the noise free simulated target dynamics in Fig. \ref{fig:noise_control} $\mathbf{C}$. The results show that higher additive noise, which influences the power system though the controller, leads to higher spectral errors. This effect increases with higher control parameters. A comparison of the controlled voltage dynamics in Figs. \ref{fig:noise14} and \ref{fig:noise12}, featuring different noise levels ($\sigma=0.012$ and $\sigma=0.014$) and control parameter values ($K=30$ and $K=10$), provides a clear visualization of the noise amplification properties of a higher control parameter. The frequency spectra of the dynamics provide the insight that large parts of the observed additional noise is located in the higher frequencies domains above $2Hz$. These domains represent the frequency ranges associated with the system's time step size, $\Delta t=0.05$, with  $f=1 / \Delta t = 20$, as the control force is applied at each time step, introducing noise to the system. The spectral errors of the control parameters for different noise levels below $2Hz$ are displayed in Fig. \ref{fig:noise_control} $\mathbf{D}$. The results show that in the lower frequency domain the amplification effect of the control parameter nearly vanishes and that the power system is accurately controlled to the frequencies domains of the true target dynamics when applied with noisy input data. To quantify the controller induced noise to the power system we calculate the signal to control noise ratio ($ScNR$) of the controlled voltage dynamics at the target state. We set the voltage of the control dynamic from NGRC as the signal and the differences between the signal and the voltage dynamic of the controlled power system as the control noise. For $K=10$ and $\sigma=0.014$ the $ScNR \approx 14.02$, which means the power of the signal $P_{signal} = 10^{14.02/10}P_{nosie}$ is approximately $25$ times larger than the control noise power. For $K=30$ and $\sigma=0.012$, the $ScNR \approx 5.32$, which shows an approximately $3.4$ times larger signal power. In Fig. \ref{fig:noise_analyis_control} $\mathbf{B}$ we compare the additive noise levels $\sigma$ with the standard deviation of the control noise $\sigma_c$. We find that for small control parameters up to $K=20$, the induced control noises to the power system $\sigma_c$ is smaller than the additive noise levels. For $\sigma=0.014$ with $K=10$ the induced $\sigma_c$ is approximately $50\%$ smaller. We observe a linear relationships between $\sigma$ in the measured dynamics $\tilde{\mathbf{u}}(t)$ and the control noise $\sigma_c$ under different control parameters $K$. The control force $\tilde{\mathbf{F}}(t) = K(\tilde{\mathbf{u}}(t) - \mathbf{v}(t))$ decomposes into $\tilde{\mathbf{F}}(t) = \mathbf{F}(t) + \mathbf{F}_n$, where $\mathbf{F}_n = K\mathbf{\mu}(t)$. When synchronisation occurs, $\mathbf{F}(t)$ vanishes, leaving only $\mathbf{F}_n$. Looking at the standard deviation of $ \mathbf{F}_n$ and integrating over the time step size $\Delta t$ following Eq. \ref{eq:control_integration}, yields $\sigma_c = K\Delta t\sigma$, which matches the observed slopes for $\Delta t=0.05$. These results show that for $K>1/\Delta t$ the control noise is amplified, while for smaller control parameters the control noise is damped.


\newpage
\subsection*{Discussion}
In this work, we introduce a model-free and data-driven machine learning methodology to control dynamical systems in previously unseen target states, with a prediction evaluation and selection scheme enabling complex control strategies. Using parameter-aware next-generation reservoir computing, we demonstrate the ability to utilize its extrapolation capabilities to efficiently generate control scenarios that are selected when a predefined control objective is met. We show that this way transient behavior can be suppressed across transition times in a non-stationary control application. Furthermore, we demonstrate that the method can navigate transitions even in scenarios where system collapse is observed frequently, while ensuring  fast transitions and avoiding prolonged transient behavior. These scenarios are validated for additive noise on the dynamics of the power system. NGRCs stand out for their exceptional data efficiency. Our experiments confirm this by requiring more than an order of magnitude less training data than results reported for learning the Lorenz system using traditional reservoir computers \cite{kim2021teaching}. This makes the approach particularly suitable for scenarios where data is scarce or expensive to collect. Despite this efficiency, NGRCs require careful hyperparameter optimization to avoid stability issues during prediction. This sensitivity is greater than in other reservoir computing approaches mainly because of these following three factors in combination: the lack of inherent boundedness can cause instability, the feature vector dimension scales combinatorially with time delays and monomial order, and higher dimensionality often fails to improve performance. Once a suitable configuration is achieved, however, the parameter-aware NGRC can generate stable non-stationary predictions and be used as a controller over transients to unseen parameter regions. Recent work, demonstrated that NGRC-based control of dynamical systems is feasible on platforms such as FPGAs \cite{kent2024controlling}. Our method extends this work through parameter awareness, enabling complex control strategies and arbitrary target states beyond the learned parameter range while maintaining equivalent computational requirements within a single trained NGRC architecture. This efficiency, combined with low training data requirements, makes the approach especially suited for edge applications where real-time adaptability is key.
Future work will focus on the sensitivity issue by refining the reservoir and its optimization algorithm to further progress its adaptability while maintaining high data efficiency with architectural simplicity. By addressing these challenges, the proposed methodology shows promise as an innovative tool for controlling dynamical systems in complex scenarios and edge environments, opening pathways to applications where traditional methods fall short.

\section*{Methods}

\subsection*{Next-Generation Reservoir Computing} Next-generation reservoir computing (NGRC) is a machine learning framework that enhances traditional reservoir computing by leveraging deterministic structures derived from non-linear vector autoregression (NVAR). Instead of utilizing a randomly weighted internal network, NGRC constructs its feature vectors through a library of unique monomials of time-shifted input variables from the input data. This process involves concatenating $k$ past time series data points, each separated by $s$ time steps, to form a linear time-delay state vector. The final feature vector $\mathbf{r}_i$ is created by generating unique monomials of orders $O$ from this state vector, introducing non-linearity into the model \cite{gauthier2021next}.
In this way, given a time series of dimensions d $\mathbf{X}$, a matrix of feature vectors $\mathbf{R}(\mathbf{X},k,s,O) = (\mathbf{r}_1,\mathbf{r}_2,\ldots,\mathbf{r}_n)$ can be generated. Training is simplified by optimizing only the output layer $\mathbf{W}_{out}$ that maps the feature vectors onto the difference between the current point and the next point in time using ridge regression, such that   
\begin{equation}
\mathbf{x}_{i+1}=\mathbf{x}_i+\mathbf{W}_{out} \mathbf{r}_{i+1}.
    \label{eq:differenceNG}
\end{equation}
To parameterize the NGRC feature vector, the approach introduced in \cite{koglmayr2024extrapolating} is followed, where the parameter $\theta$ multiplied by a scaling factor $\gamma$ is added to each element of the feature vector, resulting in $\mathbf{r}_i' = \mathbf{r}_i + \gamma\theta$. Furthermore, $\mathbf{r}_i'$ is expanded by including its own powers up to order $O_{\text{states}}$ to generate the parameterized feature vector $\widetilde{\mathbf{r}}_i$. To employ this architecture for multiple stationary dynamics during training, the concept of multifunctionality is leveraged \cite{flynn2022exploring}. Therefore, for each time series $\mathbf{X}_m$ and its parameter $\theta_m$, the parametrized feature vectors $\widetilde{\mathbf{R}}_m = (\widetilde{\mathbf{r}}_{m,1}, \widetilde{\mathbf{r}}_{m,2}, \ldots, \widetilde{\mathbf{r}}_{m,n})$ are generated along with their corresponding training targets $\mathbf{Y}_m$. 
To optimize the output layer with this setup, the parameterized feature matrices $\widetilde{\mathbf{R}}_m$ and the training target matrices $\mathbf{Y}_m$ are concatenated into single matrices $\widetilde{\mathbf{R}}_M$ and $\mathbf{Y}_M$, respectively. This allows the output layer $\mathbf{W}_{\text{out}}$ to be learned via ridge regression by solving the optimization
\begin{equation}
\mathbf{W}_{\text{out}} = \mathbf{Y}_M \widetilde{\mathbf{R}}_M^T \left( \widetilde{\mathbf{R}}_M \widetilde{\mathbf{R}}_M^T + \beta \mathbf{I} \right)^{-1},
\label{multitrain}
\end{equation}
where $\mathbf{I}$ is the identity matrix and $\beta$ is the regularization parameter. If the training is successful, the structure allows the parameter $\theta$ to change during the prediction, so that the prediction is performed under $\theta_i$
\begin{equation}
\mathbf{x}_{i+1} = \mathbf{x}_i + \mathbf{W}_{\text{out}} \widetilde{\mathbf{r}}_{i+1}(\theta_{i+1}).
\label{non-stationary equation}
\end{equation}
In this way, both stationary and non-stationary dynamics can be predicted, even for parameter ranges that were not seen during training.

\subsection*{Control}
Control theory refers to the study and application of methods for influencing the behavior of dynamical systems such that a desired target state is achieved. In the context of chaotic systems, these approaches are able to stabilize unstable periodic orbits by applying small perturbations through external forces. Traditional methods, such as OGY control \cite{ott1990controlling} and delayed feedback control \cite{pyragas1992continuous}, typically rely on phase space techniques, which require detailed knowledge of the system's underlying equations or extensive datasets. These approaches are typically limited to steering systems toward simple dynamical target states. Although various extensions have been proposed \cite{boccaletti2000control}, including methods to "chaotify" periodic or synchronized dynamics \cite{schiff1994controlling}, the ability to reach arbitrary dynamical target states has only recently been demonstrated \cite{haluszczynski2021controlling}. In further research \cite{haluszczynski2023controlling}, this novel approach was used to compare traditional reservoir computing with next-generation reservoir computing (NGRC) in terms of data requirements for effective control. NGRC demonstrated superior performance, requiring ten times less training data while achieving comparable control capabilities.

In this paper, we build on the control mechanism introduced in \cite{haluszczynski2021controlling,haluszczynski2023controlling} to guide dynamical systems into previously unseen target states. This general framework is not restricted to the next-generation reservoir computer employed here and can also be implemented with alternative methods, such as the minimal reservoir computer, traditional reservoir computing techniques, or other suitable machine learning techniques. The core idea is to train the machine learning method to predict the desired dynamics of the system, represented as the state $\textbf{X}(\rho_i)$ at a given system parameter $\rho_i$. In particular, the dynamics of the desired state $\textbf{X}(\rho_i)$ does not need to be recorded prior to training. Instead, the system can leverage the ability to predict unseen states, as explained in the previous Section, and be trained using the available data. If the system instead evolves to a different state $\textbf{Y}$ --- due to external influences or changes in its bifurcation parameters $\rho_j$ --- an external force $\textbf{F}(t)$ is applied to guide the dynamics back to the desired state $\textbf{X}(\rho_i)$. Without this intervention, the system would persist in state $\textbf{Y}$, following its natural trajectory $\textbf{u}(t)$.

To compute the appropriate control force $\textbf{F}(t)$, it is necessary to predict how the trajectory of the system $\textbf{u}(t)$ would evolve if it were in the desired state $\textbf{X}(\rho_i)$. This hypothetical trajectory is defined as $\textbf{v}(t)$. Assuming that the reservoir computer has been trained effectively, it can accurately forecast the future trajectory $\textbf{v}(t)$ as if the system were in state $\textbf{X}(\rho_i)$. By comparing this predicted trajectory with the actual state evolution of the system $\textbf{Y}$, the control force is calculated based on the difference between the actual and hypothetical trajectories:
\begin{eqnarray}
\textbf{F}(t)=K(\textbf{u}(t) - \textbf{v}(t)) \ ,
\label{eq:force}
\end{eqnarray} 
where $K$ is a control parameter. This control force depends only on the observed actual system coordinates and the predicted coordinates, which requires no knowledge of the system's underlying equations. This feature makes the approach highly versatile and applicable to real-world problems where precise mathematical models are unavailable. For a thorough analysis, we demonstrate this control strategy on mathematical example systems simulated using its known equations instead of observing a real system. In this case, the control force is incorporated by adding it to the system's differential equations $\dot{f}$. The updated equations are then solved at each time step, incorporating the control force into the system's evolution 
\begin{eqnarray}
\textbf{u}(t+\Delta{t}) = \int_{t}^{t+\Delta{t}} (\dot{f}(\textbf{u}(\tilde{t})) + \textbf{F}(\tilde{t})) d\tilde{t} \ .
\label{eq:control_integration}
\end{eqnarray} 
To ensure a smooth transition in Lorenz application to the system's target state without transients occurring, a prediction of the target state must be selected that does not exhibit transient behavior. Multiple prediction realizations are generated for a given target state and their correlation dimension is calculated. If the correlation dimension of a prediction is below the rcd threshold value, where $\text{rcd} < 1.21$  which represents the highest correlation dimension of the stationary target dynamics, it is considered free from transients and can be used for control. Depending on the reservoir computing method used, various strategies can generate multiple prediction realizations for a given target state. In this approach, we varied the initial conditions slightly and selected the transient-free prediction that starts closest to the system's current coordinates when applying the control force. For the power system application with noise, the control method is identical, except for $ \mathbf{u(t)} \rightarrow\tilde{\mathbf{u}}(t) = \mathbf{u}(t) + \mathbf{\mu}(t)$ with $\mathbf{\mu}(t)  \sim \mathbf{\mathcal{N}}(0,\,\sigma^{2})$. The resulting control force is applied to the dynamical system directly, as well.

\subsection*{Lorenz System}

We selected the Lorenz system \cite{lorenz1963deterministic} as an example to demonstrate the functionality of the control mechanisms. In the range of bifurcation parameters analyzed of $\rho$, the Lorenz system exhibits periodic behavior for $\rho > 100$, undergoes period doubling for $\rho < 100$ and transitions to chaotic behavior.
The system of equations $\dot{f}(\mathbf{x}(t,\rho))$ is given by
\begin{equation}
    \dot{x} = \sigma (y - x), \\
    \dot{y} = x (\rho - z) - y, \\
    \dot{z} = x y - \beta z,
\end{equation}
where the parameters are set to $\sigma = 10$ and $\beta = \frac{8}{3}$. These equations are solved for the simulation results using the fourth-order Runge-Kutta method with a time-step size of $\Delta t = 0.005$.
The results were obtained using an NGRC architecture with parameters $k = 2$, $s = 47$, $O = [1, 2, 3]$, $O_{\text{states}}= 3$ and $\gamma = 0.18$. In addition, a bias feature of $1$ was included in the parameterized feature vector. During the control, the control parameter was set to $K=50$ across all transition periods.

\subsection*{Generic power system model}

The model is defined by a system of four ordinary differential equations:

\begin{equation}
    \dot{\delta}_m = \omega,
\end{equation}
\begin{equation}
    M\dot{\omega} = -d_{m}\omega + P_m - E_mY_m\sin(\delta_m - \delta)V,
\end{equation}
\begin{equation}
    K_{qw}\dot{\delta} = -K_{qv2}V^2 - K_{qv}V + Q(\delta_m, \delta, V) - Q_0 - Q_1,
\end{equation}
\begin{align}
    TK_{qw}K_{pv}\dot{V} = &\; K_{pw}K_{qv2}V^2 + (K_{pw}K_{qv} - K_{qw}K_{pv})V \nonumber \\
    &+ K_{qw}[P(\delta_m, \delta, V) - P_0 - P_1] \nonumber \\
    &- K_{pw}[Q(\delta_m, \delta, V) - Q_0 - Q_1]
\end{align}
The real power demand $ P$ and reactive power demand $Q$ are defined as:

\[
\begin{alignedat}{2}
  P(\delta_m,\delta,V) &= -E'_0Y'_0V\sin\delta
                          + E_mY_mV\sin(\delta_m-\delta)
  &\qquad
  Q(\delta_m,\delta,V) &= -E'_0Y'_0V\cos\delta
                          - (Y'_0 + Y_m)V^{2}
                          + E_mY_mV\cos(\delta_m-\delta)
\end{alignedat}
\]
\noindent
In this formulation, the real‑power demand~$P$ and reactive‑power demand~$Q$ drive the differential equations governing the load voltage~$V$ and the motor angle~$\delta$. The term~$\delta_m$ denotes the angle separation of the rotor between the two generators, while $\omega$ represents the speed of the generator rotor. For a comprehensive technical discussion, see Dobson\,\emph{et al.}~\cite{dobson1989towards}. The parameters are selected according to the conventions used in previous reservoir computing models~\cite{kong2021machine,koglmayr2024extrapolating}.

\begin{equation}
\begin{alignedat}{3}
  E'_0 &\,= \frac{E_0}{\sqrt{1 + C^{2}Y_0^{-2} - 2CY_0^{-1}\cos\theta_0}}
  &\qquad
  Y'_0 &\,= Y_0\sqrt{1 + C^{2}Y_0^{-2} - 2CY_0^{-1}\cos\theta_0}
  &\qquad
  \theta'_0 &\,= \theta_0 + \tan^{-1}\!\bigl(
               \tfrac{CY_0^{-1}\sin\theta_0}{\,1 - CY_0^{-1}\cos\theta_0\,}
               \bigr)
\end{alignedat}
\label{eq:EYtheta}
\end{equation}
 
The parameter values are:

\[
\setlength{\jot}{4pt}               
\begin{alignedat}{8}                
  K_{pw}  &= 0.4    &\qquad&
  K_{pv}  &= 0.3    &\qquad&
  K_{qw}  &= -0.03  &\qquad&
  K_{qv}  &= -2.8   &\qquad&
  K_{qv2} &= 2.1 \\[3pt]
  T       &= 8.5    &\qquad&
  P_0     &= 0.6    &\qquad&
  P_1     &= 0      &\qquad&
  Y_0     &= 3.33   &\qquad&
  Y_m     &= 5\\[3pt]
  P_m     &= 1      &\qquad&
  d_m     &= 0.05   &\qquad&
  \theta_0&= 0      &\qquad&
  E_m     &= 1.05   &\qquad&
  M       &= 0.01464\\[3pt]
  C       &= 3.5    &\qquad&
  E_0     &= 1      &\qquad&
  Q_0     &= 1.3
\end{alignedat}
\]

\noindent
The parameter \(Q_1\) acts as the bifurcation parameter, representing the reactive power demand of the system.  
The bifurcation diagram is produced by integrating the governing equations with a fourth‑order Runge–Kutta (RK4) scheme.  
Each simulation begins from the initial state
\[
\mathbf{x}_0 = (\delta_{m,0}, \omega_0, \delta_0, V_0)^{\mathsf T}
             = (0.17,\,0.05,\,0.05,\,0.83)^{\mathsf T},
\]
and is advanced for \(20\,000\) time steps with a fixed step size of \(\Delta t = 0.05\).  
The bifurcation parameter is incremented by \(\Delta Q_1 = 2 \times 10^{-6}\) in the analysis of the bifurcation diagrams.
The training data are obtained for the following values of the bifurcation parameter:
\[
Q_1 \in \{2.98947,\,2.98952,\,2.98957,\,2.98962,\,2.98967,\,2.98972,\,2.98977\}.
\]

\noindent
The hyperparameters of the trained NGRC are $k = 2$, $s = 63$, $O = [1, 2]$, $O_{\text{states}}= 2$ with an additional bias term, and $\gamma = 0.292$.
\noindent
For the non-stationary predictions of the instantaneous parameter change, we hold $Q^{initial}_1$ for 250 seconds constant and then switch to $Q^{target}_1$ for additional 500 seconds. 

\subsection*{Prediction evaluation and selection scheme}

Control scenarios are predicted with Eq.~\ref{non-stationary equation}.
The vector $\boldsymbol{\theta}$ specifies the desired non‑stationary dynamics, and its length sets the length of the prediction of the NGRC. For each element in $\mathbf{\theta}$, the NGRC predicts the next point in time under the current $\theta_i$. Varying the NGRC warm‑up phase yields distinct non‑stationary trajectories. Each trajectory is evaluated against a predefined objective- for example, the running correlation dimension of the trajectory must remain below a threshold or exhibiting short transition periods measured by the spectral entropy over time. 
A trajectory that meets the objective is adopted as the control strategy.  
In the applications discussed, we start from $\theta_{initial}$ for some time to control the system from its initial state to the initial state of the prediction before the parameter change is forced.

\subsection*{Running window correlation dimension}
To determine whether the system undergoes a smooth transition during the non-stationary process, we employ a running correlation dimension approach computed over a defined window length. This technique captures the dynamic spatial complexity of the system over time. For periodic dynamics, the running correlation dimension values remain close to 1, whereas they exceed 1 when the dynamics exhibit chaotic behavior. In the case of the chaotic target dynamics of the example Lorenz system with $\rho = 99.5$, this value remains below $1.21$. During non-stationary simulation, prediction, and control, we detect a non-smooth transition if the running correlation dimension surpasses $1.21$, indicating possible transients, while a smooth transition is considered if it remains below this threshold. For each windowed trajectory, the correlation dimension $\nu$ quantifies the fractal dimensionality of the trajectory's phase space \cite{grassberger2004measuring}
\begin{eqnarray}
C(r) &= {\frac{1}{W^2}\sum^{W}_{i,j=1}\theta(r- | \textbf{x}_{i} - \textbf{x}_{j}  |)}
\label{eq:corrintegral}
\end{eqnarray}
where $\theta$ is the Heaviside function. The states are considered close to each other if their distance is less than the threshold distance $r$. The dimension $\nu$ is determined from the scaling relation
\begin{eqnarray}
C(r) \propto r^{\nu}.
\label{eq:corrdim}
\end{eqnarray}
We compute $\nu$ using the Grassberger–Procaccia algorithm \cite{grassberger1983generalized} and use a window length of $W = 500$ time steps for the Lorenz system.

\subsection*{Spectral Entropy Measure}\label{sec:spec_entropy}

The spectral entropy metric is used to detect intermittent episodes in the non‑stationary power‑system model, capturing the transition from one periodic regime to another.
We evaluate the short-time Fourier transform \cite{2020SciPy-NMeth} (STFT) calculated with a window of \(n_{\text{perseg}} = 2000\) samples and an overlap of \(n_{\text{overlap}} = 1950\) samples, i.e. 97. 5\%. Only components in the \(0.1\text{–}2~\text{Hz}\) band are retained for subsequent processing, as this range is known to carry the dynamics of interest in the system analyzed. 
\noindent
Frames whose center times violate $n_{\text{perseg}}$ at the start and end of the analyzed time series are discarded. For every remaining STFT frame \(\ell\) the band‑limited power spectrum
\(P_{k\ell} = |Z_{xx}(f_k,t_\ell)|^{2}\) is normalized to the unit mass and its
Shannon entropy\cite{shannon1948mathematical} is evaluated,
\begin{equation}
  H_\ell \;=\;
  -\sum_{k} P_{k\ell}\,\ln P_{k\ell},
  \label{eq:entropy}
\end{equation}
yields the trace of entropy \(H_\ell\) against the center time~\(t_\ell\).
A first-order difference
\(\Delta H_\ell = H_\ell - H_{\ell-1}\) is used as a discrete approximation of the temporal derivative. The onset of an intermittent episode is flagged at the earliest time index for which \(\Delta H_\ell\) exceeds a fixed rise threshold \(\Delta H_{\text{rise}} = +0.01\).
The episode is considered to end with the first subsequent index where \(\Delta H_\ell\) falls below the corresponding fall threshold \(\Delta H_{\text{fall}} = -0.01\). The duration of the detected intermittent period is adjusted to match the number of segments used in the analysis.

\section*{Author contributions statement}
D. K. and A. H. developed the control algorithm and performed the experimental studies. C. R. initiated and supervised the work. All authors interpreted the findings and wrote the manuscript.

\section*{Data and Code Availability}

The data and code are available from the corresponding author upon reasonable request.

\section*{Acknowledgments}
D. K. gratefully acknowledges the funding provided by Allianz Global Investors.

\section*{Declaration of Interests}
The German Aerospace Center, together with A.H., have filed a patent application related to this work.

\bibliography{sample}

\end{document}